\documentclass[runningheads,a4paper]{llncs}

\usepackage{amssymb}

\usepackage{graphicx}

\usepackage[usenames,dvipsnames]{xcolor}
\definecolor{lightgray}{gray}{0.95}

\usepackage{listings}
\usepackage{url}
\urldef{\mailsa}\path|{machens, turau}@tuhh.de|    
\newcommand{\keywords}[1]{\par\addvspace\baselineskip
\noindent\keywordname\enspace\ignorespaces#1}

\begin{document}

\mainmatter  

\title{Opacity of Memory Management in Software Transactional Memory}

\titlerunning{Opacity of MM in TM}

\author{Holger Machens\and Volker Turau}
\authorrunning{Opacity of MM in TM}

\institute{Institute of Telematics\\
Hamburg University of Technology\\
\mailsa\\
\url{http://www.ti5.tuhh.de}}

\toctitle{Opacity of Memory Management in Transactional Memory}
\tocauthor{Opacity of MM in TM}
\maketitle

\begin{abstract}
Opacity of Transactional Memory is proposed to be established by incremental
validation. Quiescence in terms of epoch-based memory reclamation is applied to
deal with doomed transactions causing memory access violations. This method
unfortunately involves increased memory consumption and does not cover
reclamations outside of transactions. This paper introduces a different method which combines incremental validation with elements of
sandboxing to solve these issues.
\keywords{transactional memory, opacity, privatization, memory reclamation}
\end{abstract}

\section{Introduction}
The stagnating improvement in processing speed and the increasing availability
of multi- and many-core processors have led to a growing interest in
methods and tools to ease concurrent and parallel programming. Transactional
Memory (TM) is considered as a promising candidate to replace traditional mutual
exclusion in critical sections by more intelligent concurrency control (CC)
methods. TM eliminates deadlocks and generally improves the
scalability on multi-core machines allowing more concurrency in critical sections.

As the name suggests, TM provides transparent use of transactions on shared data
in main memory. Transactions, as known from database or distributed systems,
speculatively execute a set of instructions to be rolled back in case
of data inconsistency or deadlock. A rollback consists of aborting the
execution, a discard of all modifications and a restart at the beginning. 

Especially when performing dirty reads there is a critical time period between
the occurrence of a data inconsistency and its detection. In this period the
transaction works on inconsistent data and may run in multiple different
problems: Pointers can be invalid, expressions in conditional branches may
have wrong results and lead to endless loops or parameters to operating
system resource allocations (such as memory allocation) may be too big,
resulting in resource exhaustion. Transactions which run in such problems are
known as \textit{doomed transactions} and \textit{opacity} is the property of TM
systems to hide the side effects of doomed transactions.

Incremental validation is a known method to achieve opacity in TM systems
written in software (Software Transactional Memory, STM). Accordingly, on each
read of data a transaction validates all its previously read data by a
comparison of a formerly taken copy to the current state of the data in memory.
This method is save in regards to modifications in memory, but not in
respect to memory reclamation (freeing memory) which can cause unexpected
memory access violations in transactions and process termination as its
consequence.

Hudson et al. presented an algorithm for memory management (memory allocator)
for STM which prevents the occurrence of segmentation
faults caused by inconsistencies in doomed transactions. Main concept of the
allocator is to buffer and defer reclamations of memory blocks
and perform them when no transaction exists which could eventually try to
access the affected memory blocks.

Obviously, deferring memory reclamation has some disadvantages such as increased
memory consumption. Thus, we have analyzed the side effects of this method and
possible solutions to reduce or prevent them. Based on this knowledge we have
developed a different method, which significantly reduces memory consumption and
slightly improves the response time of the allocator functions (allocation and
reclamation). The results have been evaluated by comparison using benchmarks
such as the Stanford Transactional Applications for Multi-Processing (STAMP).

The following section will give a brief introduction in the problem of
opacity in TM and existing solutions (Section
\ref{sec:stateoftheart}). Section \ref{sec:opaqueMM} then discusses the memory
allocator of Hudson et al. and its side effects. In Section
\ref{sec:design} the design of our method is explained and in Section
\ref{sec:evaluation} the evaluation and results are presented. In the end we
give a short conclusion and prospective on future work.

\section{Opacity in Software Transactional Memory}
\label{sec:stateoftheart}
Explaining doomed transactions and opacity \cite{bib:OPACITY} requires at least
a simplified model of transactions in STM: In a TM-enabled application certain
critical code sections are marked to be executed as a transaction by the
currently active thread.  Those sections are typically instrumented with calls
into the STM runtime library, which enforces the concurrency control.  A
transaction performs an atomic transformation of shared data from one consistent
state into another.  Thus, the shared data is inconsistent if at least one
transaction started to update shared data and has not yet finished.  In turn,
the state of a transaction is said to be valid if all data read by it (its
\textit{read-set}) originates from exactly one consistent state.

Currently, the most efficient STM implementations are based on optimistic
concurrency control algorithms. Transactions are executed speculatively assuming
that there are no concurrent transactions in the first place. Thus, those
transactions will inevitably run into inconsistent states that are meant to be
detected and solved later by a rollback. During the time period between
occurrence and detection the transaction works on inconsistent data and produces
unexpected side effects. While some side effects are of minor relevance others
can crash the application and even the entire operating system. A TM system
guarantees opacity if it hides all these side effects. 

This is an extended list of possible side effects originating from
\cite{bib:SANDBOXING}:

\begin{itemize}
  \item Runtime errors
  \begin{itemize}
    \item Segmentation faults, caused by erroneous pointers
    \item Arithmetic faults, caused by erroneous operands (such as null divisor)
    \item Illegal instruction faults, caused by erroneous code modifications
    (considering self modifying code) or call or jump targets
  \end{itemize}
  \item Bypassing concurrency control due to erroneous call or jump targets
  \begin{itemize}
    \item Execution of non-transactional code 
    \item Exit without commit
  \end{itemize}
  \item Non-terminating code, caused by invalid values in branch or abort
  conditions
  \begin{itemize}
    \item Infinite loops
    \item Infinite recursions
  \end{itemize}
  \item Resource exhaustion, caused by invalid parameters to resource allocation
  functions, such as memory allocation
\end{itemize}

While the reason for most of the errors above arise directly from
the value received by a dirty read others occur as a consecutive error to
unintended modifications to data on the stack (e.g. when crossing the
boundaries of a local array). Most error prone are stack pointer (SP) and base
pointer (BP) which are used to access local data via an offset and the return
address which points to the instruction to return to from the current function.
As soon as SP or BP gets invalid, every access to local data is invalid as well
which leads to any of the errors above.

There are mainly three different approaches to prevent those errors which will
be explained in the following sub-sections.

\subsection{Mutual Exclusion}
The easiest way to provide opacity is to use locks to guarantee mutual exclusion
on shared data according to a pessimistic concurrency control approach. Thus,
dirty reads are forbidden per definition and doomed transactions cannot occur.
But use of locks in TM has several disadvantages:
\begin{itemize}
  \item In native programming languages such as C locks have to be associated
  with memory blocks of certain size (granularity). Lacking alternatives, locks
  have to be stored in a global table separated from the memory blocks.
  Entries in this table are called owner records (\textit{orecs}), because they
  were originally introduced to store ownership of objects. Thus, each
  access to a memory 'object' requires an additional access to the orec which 
  in turn increases cache miss rate on hardware level and thereby reduces
  scalability of the TM.
  \item Locking protocols involve waiting time which is in some cases just
  wasted. Consider a transaction $A$ which waits on a lock held by transaction
  $B$. Transaction $B$ runs in a deadlock and performs a rollback, thereby
  releasing the lock. Thus, $A$ has been waiting for nothing. Additionally,
  those wait-for relationships are transitive, which means, another transaction may have been
  waiting for $A$ as well and the case can occur again after rollback.
  Considering at least one transaction to successfully finish in a conflicting situation the worst case waiting time for the single
  transaction still grows quadratically with the number of active transactions.
  \item Increasing the granularity of memory objects associated with locks
  reduces the probability of conflicts, cache misses and waiting time in turn
  but it reduces the concurrency and scalability as well.
\end{itemize}

Another method to reduce the waiting time is to allow so-called lock stealing,
where a transaction may decide to steal the lock currently held by another
transaction and proceeds with its work without waiting (cf. \cite{bib:SKYSTM}).
But this eliminates the guaranteed mutual exclusion and again requires strategies to deal with
doomed transactions as applied for optimistic concurrency control mechanisms
explained in the next sections.

\subsection{Sandboxing}
A way to achieve opacity in optimistic concurrency control is to embed
transactions in a sandbox which prevents some errors of doomed transactions and
transparently handles the remaining errors. This approach is almost similar to
the execution of intermediate code in a java virtual machine or a common
language runtime.

The most comprehensive existing example for sandboxing in C has been published
by Dalessandro and Scott \cite{bib:SANDBOXING}. They applied sandboxing on an
STM implementation which uses orec-based validation and deferred updates. That
means in particular, that every write access to any application data is just
stored in a local log (so-called \textit{write-set} or \textit{redo-log}) of the
transaction. The actual write to the shared data is performed after the
transaction has been checked to be valid at commit time.

The proposed sandbox additionally provides the following mechanisms to guarantee
opacity of transactions:

\begin{itemize}
  \item Catching runtime errors in signal handlers which validate and abort the
  transaction in case of inconsistency.
  \item Timer triggered repeated validation to escape from endless loops or
  recursions.
  \item Validation prior to the execution of indirect jumps or function
  calls inside transactions to prevent bypassing of the CC mechanism. 
  \item Validation of the parameters to memory allocation functions to prevent
  memory exhaustion.
  \item Validation of potential access to SP.
\end{itemize}

Hence, besides the deferred update for any data, the main method applied to
achieve opacity is to insert validation at critical points in the transactional
section by instrumentation. Sandboxing is said to cause more latency but
Dallessandro and Scott have shown good performance and scalability of their
approach at least with the benchmarks they have used. Others criticize
sandboxing for potentially overriding signal handlers of the application but this can be
solved by chaining the signal handlers as it was illustrated in their work as
well.

\subsection{Incremental Validation}

A widely accepted method to ensure opacity is incremental validation as proposed
in \cite{bib:OPACITY}. The fundamental principal here is to validate the whole
read-set on each read (cf. Listing \ref{lst:txread}). Thus, a validation
consists of a validation of every entry in the read-set on each consecutive
read. This validation can be performed according to two methods:
\begin{enumerate}
  \item \textit{value-based}: The read-set contains the value first read
  by the transaction which is compared against the current content of the
  originating location in memory. Differences indicate inconsistency (see for
  example NOrec \cite{bib:NOREC}).
  \item \textit{orec-based}: The read-set contains the value and the state of
  the orec seen on the first read access. The orec state might be
  represented by a version number for example, which is incremented on each
  write access to the associated location in memory. In case of lock stealing, 
  the orec will contain the current value of the lock (e.g. lock owner, see for
  example SkySTM \cite{bib:SKYSTM}).  Instead of accessing the originating
  location in memory this method just validates against the state of the associated orec comparing either the
  version number or lock state. Version differences or lost locks indicate inconsistencies in
  this case.
\end{enumerate}

\begin{center}
\begin{minipage}{0.75\textwidth}
\begin{lstlisting}[caption={Simple read function with incremental
validation},label=lst:txread]
int txread(int* addr) {
	if (!%*\textbf{\textit{validate()}}*)) rollback_and_restart(); 
	int val = *addr;
	if (!%*\textbf{\textit{validate()}}*)) rollback_and_restart(); 
	append_to_readset(addr);
	return val;
}
\end{lstlisting}
\end{minipage}
\end{center}

Due to the repeated validation of the whole read-set on each transactional read,
the runtime complexity of incremental validation rises quadratically with the
number of reads. The average effort can be reduced using for example a global
commit counter, which is incremented with each commit of transactions. Thus, a
transaction can skip validations as long as the commit counter has not been
incremented. But the worst case complexity of this method is still quadratically
rising with the number of reads considering enough concurrent transactions.
Another issue of a global commit counter or orec-based validation is that they
cannot deal with concurrent modifications on shared data by concurrent threads
which do not run in a transaction because they simply ignore this mechanisms.
Thus, the developer has to make sure that those cases will not occur.

Considering the runtime complexity there is a trade-off between sandboxing and
incremental validation. While incremental validation applies to each
read of shared data sandboxing validates critical instructions only, but the
amount of critical instructions can be higher then the amount of shared
reads. Thus, both approaches have to be further analysed in future work and
maybe combined to cover both cases.

\section{Quiescence for Memory Reclamations}
\label{sec:opaqueMM}

A remaining problem of incremental validation arises through memory reclamation
(freeing memory) in transactions. Those STMs that support memory management in
transactions generally implement the following basic algorithm:
\begin{itemize}
  \item Memory allocations are performed directly and stored
  in a log simultaneously. In case of a rollback all the logged memory
  allocations have to be returned to the memory management (i.e. freed).
  \item Memory reclamations just get stored in a log. 
  They will be executed if the transaction commits or they will be discarded if
  the transaction rolls back.
\end{itemize}

Considering properly implemented concurrent applications a memory block (or
references on it) will be privatized \cite{bib:PRIVATIZATION} before it gets
freed. That means, that a shared pointer referencing this memory block will be
altered to indicate that it is no longer valid. This might be achieved by
assigning NULL to the pointer  (see Listing \ref{lst:privatization}) or removing
it from a list for example. Before accessing the data other threads must test
the reference first (see for example Listing \ref{lst:tx.ptr.access}). A
thread, which does not follow this protocol will inevitably run into a segmentation fault when the associated memory is freed.
Thus, we do not need to consider applications which do not perform a
privatization prior to a reclamation of a shared memory block.

\begin{center}
\begin{minipage}{0.7\textwidth}
\begin{lstlisting}[caption={Example of privatization},label=lst:privatization] 
/* privatize ptr */ 
void* local_ptr = shared_ptr;
shared_ptr = NULL;
 
/* reclamation */
free(local_ptr);
\end{lstlisting}
\end{minipage}
\end{center}

A transaction trying to access a probably privatized pointer is depicted in
Listing \ref{lst:tx.ptr.access}. In Line 2 the transaction reads and validates
the shared pointer (\texttt{shared\_ptr}) which is probably privatized using the
function in Listing \ref{lst:txread}. In Line 3 it makes sure that the pointer
is not privatized before it accesses the referenced memory location in Line 4.

\begin{center}
\begin{minipage}{0.75\textwidth}
\begin{lstlisting}[caption={Transaction accessing the privatized pointer
ptr},label=lst:tx.ptr.access] 
transaction {
	int* ptr = txread(shared_ptr);
	if (ptr != NULL) {
		int val = txread(*ptr);
		/* ... */
	}
}
\end{lstlisting}
\end{minipage}
\end{center}

The following cases may occur in respect to memory reclamations:
\begin{enumerate}
  \item The pointer on a memory block is privatized and freed in a transaction.
  \item The pointer on a memory block is privatized in the transaction but freed afterwards
  without running a transaction.
  \item The pointer on a memory block is privatized and freed without running a
  transaction.
\end{enumerate}

Even with incremental validation the first two cases already cause a problem:
After the last validation of the transaction (line 2 in Listing
\ref{lst:txread}) and the actual access to the memory location (Line 3 in
Listing \ref{lst:txread}) the privatizing transaction could perform its commit
and free the memory. Thus, the accessing transaction will receive a segmentation
fault (SkySTM for example suffers this problem in case of lock stealing).
This can be prevented using a global lock to establish mutual exclusion between
concurrent commits and reads. But this would heavily decrease the concurrency
proportional to the number of reads during transaction execution.

A method which solves the privatization problem without mutual exclusion is
called \textit{quiescence} mechanism \cite{bib:QUIESCENCE}. The general idea is to
defer access to privatized data until every active transaction has noticed the
privatization. One implementation of this concept is to perform the
privatization and then block the privatizing transaction before it modifies the
privatized data until every concurrent transaction has either committed or
aborted. Thus, all remaining active transactions will read the new value of the
privatized data (pointer in our case). That means, there is a time when
quiescence on that privatized data has been achieved and the privatizing
transaction can safely proceed (free the associated memory in our case).
Disadvantage of this method is, that it involves waiting for other transactions which might be
of arbitrary duration.

In case of incremental validation the described problem arises from memory
reclamations only. All the other data inconsistencies in respect to
privatizations will be detected through the validation after the dirty read
(i.e. in Line 4 of Listing \ref{lst:txread}). Thus, it is enough to defer just
memory reclamations to a time of quiescence. This is for example implemented in
NOrec as a so-called epoch-based memory reclamation
\cite{bib:ALLOCATOR,bib:EPOCHRECLAMATION}: The execution time of a concurrent
application gets logically partitioned into so-called global epochs. The first
global epoch begins at the start of the application. The lifetime of each thread
is partitioned in thread-specific epochs as well. A new epoch of a thread begins
with the start of the thread and every start, restart and end of a transaction.
A new global epoch begins each time when all threads have switched into a new
thread-specific epoch. Every memory reclamation request in transactions gets
associated with the currently running global epoch and stored in a global data
structure called \textit{limbo}. The limbo is checked by each committing
transaction for reclamation requests that have reached the point of quiescence
(older than two global epochs). Those requests will then be executed by the
currently committing transaction.

\begin{center}
\begin{minipage}{0.7\textwidth}
\begin{lstlisting}[caption={Privatization 'without'
transaction},label=lst:txprivatize] 
/* privatize ptr */ 
void* tmp = ptr;
ptr = NULL;
 
/* transaction-aware barrier */
transaction { /* intentionally empty */ }

/* reclamation */
free(tmp);
\end{lstlisting}
\end{minipage}
\end{center}

The third case listed above generally requires to manually apply the quiescence
mechanism in some way. For example a barrier between privatization and memory
reclamation which is interoperable with the CC used in other threads accessing
the memory block. This barrier might be to temporarily acquire a lock in case of
mutual exclusion or run an empty transaction in case of STM (cf. Listing
\ref{lst:txprivatize}). In other words: the developer is responsible for the
correct outcome of the privatization when performed without a CC mechanism.


The epoch-based memory reclamation still has some disadvantages:
\begin{description}
\item[Impact on Memory Reclamation in General:] 
Memory reclamations outside of transactions usually do not consider the
quiescence mechanism. Thus, the second case described above is not covered and
can still cause memory access violations. Only the instrumentation of all
memory reclamations even those outside of any critical section would solve this
problem. This is technically possible but it has a negative impact on the
performance of the whole application.
\item[Increased Memory Consumption:]
Deferring memory reclamation obviously causes higher memory consumption as known
from garbage collection (GC) systems in managed code environments. Those systems generally know at each time whether references on certain
memory blocks still exist or not. Thus, GC could release the free memory if
required. In contrast the epoch-based memory reclamation cannot release
all vacant memory at any point in time. Thus, besides the fact that it causes
higher memory consumption in general, it will not be able to solve \textit{out
of memory} situations even if vacant memory is available in the limbo.
\end{description}

\section{Incremental Validation without Quiescence}
\label{sec:design}
Incremental validation is a simple method to guarantee opacity except for memory
access violations due to concurrent reclamations. Because of the disadvantages
of epoch-based reclamation we developed another method which combines incremental 
validation with an element of sandboxing to solve the memory access violations.
We have chosen NOrec as the basis for our STM prototype to allow direct
comparison to epoch-based memory reclamation. NOrec uses deferred updates,
incremental validation optimized by a commit counter and epoch-based memory
reclamation. Modifications to NOrec mainly affect the memory management and an enhancement to handle segmentation faults:
\begin{description}
\item[Allocator:]
Our simplified allocator replaces the epoch-based memory reclamation. It
implements the basic memory management algorithm for transactions explained in
the beginning of Section \ref{sec:opaqueMM}: Allocation and reclamation requests
are logged. The log is discarded on each rollback. Allocation requests get
executed instantly and freed in rollbacks. Reclamation requests from the log are executed on commit.

\item[Handling Memory Access Violations:] A signal handler has been introduced
to deal with segmentation faults. Default behavior of a segmentation fault of
an application is a process termination. Therefor, we optimized the case where
the handler is not called upon an application error: If the handler is called
inside an active transaction and the commit counter indicates a modification
since we have last validated, we first consider a conflict as the reason and
instantly perform a rollback. If there was really a conflict the transaction
will probably succeed in its next try. If there was no conflict the transaction
will receive the segmentation fault again but this time the commit counter is
eventually not modified which indicates a valid read-set and thus an error of
the application. In the latter case the error is escalated and the application
error gets visible.
\end{description}

This method implicitly solves even the cases where the reclamation request
occurs outside of a transaction (cf. Section
\ref{sec:opaqueMM}).

We did not implement chaining of signal handlers for our prototype but it just
requires to override the runtime library functions to install signal handlers
(e.g. \texttt{signal} and \texttt{sigaction}). Those have to be modified to keep our
signal handler in front of the chain and the signal handler has to call
the next signal handler in the chain or terminate the process to escalate
signals.

\section{Evaluation}
\label{sec:evaluation}
A comparison of \texttt{NOrec} with our modified version \texttt{NOrecSig}
allows an evaluation of the runtime performance in terms of scalability on
parallelization and memory consumption by measurements.  For orientation
purposes a third STM implementation called \texttt{CGL} has been measured which
establishes mutual exclusion between all critical sections (i.e. transactions) using a single global lock. Differences
between \texttt{CGL} and STM algorithms are known and not discussed
here.

We have selected those benchmarks of the Stanford Transactional Applications
for Multi-Processing (STAMP, \cite{bib:STAMP}) which contain memory reclamations
in transactions (except \texttt{yada} which had a bug in our version). To
demonstrate worst case behavior we implemented another benchmark called
\texttt{opacity} which aggressively repeats concurrent memory allocations and
reclamations. 

\begin{description}
\item[intruder:] Simulated intrusion detection system processing
different attacks.
\item[ssca2:] A benchmark operating on a huge directed multi-graph which
concurrently adds and removes nodes.
\item[vacation:] Simulation of a reservation system for resources such as
flights, rooms etc. with configurable amount of clients.
\item[opacity:] Mimics an application with multiple threads rapidly exchanging
messages via a shared message queue.
\end{description}

Measurements ran on a multi-core machine with four 3GHz AMD Opteron
6282SE processors, with 16 Bulldozer Cores each, a 128GB main memory,
Debian Linux with Kernel 2.6.32-5 and a GCC 4.4.5. The number of
threads has been increased up to 32. Measurements have been repeated until a
confidence interval of $5\%$ of the interval $[min,max]$ at a confidence level
of $95\%$ has been reached. Graphics with redundant information have been
removed to reduce the length of this section.

\begin{figure}[ht]
    \begin{minipage}[c]{0.5\textwidth}
    \includegraphics[width=\textwidth]{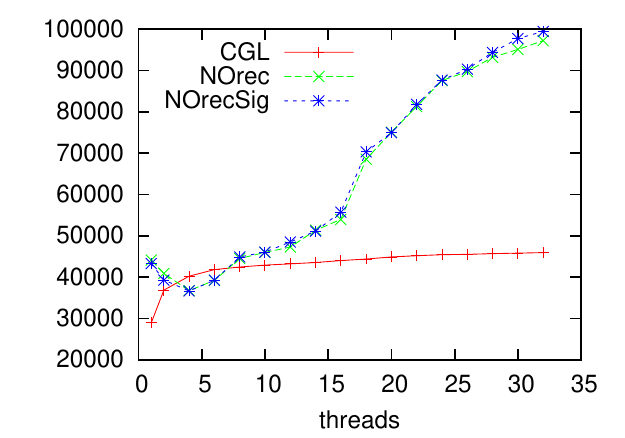}
	\caption{intruder: Exec. time [ms]}
	\label{fig:duration.intruder}
	\end{minipage}
\quad
    \begin{minipage}[c]{0.5\textwidth}
    \includegraphics[width=\textwidth]{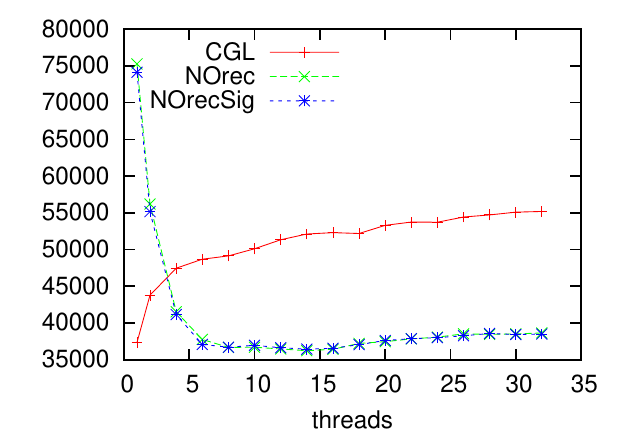}
	\caption{vacation: Exec. time [ms]}
	\label{fig:duration.vacation}
	\end{minipage}
\end{figure}

The execution time for both implementations was in all our observations almost
similar with a slight advantage of the \texttt{NOrecSig} variant (cf.
Figures \ref{fig:duration.intruder} and \ref{fig:duration.vacation}, execution
time in milliseconds). 

The memory consumption has been evaluated by detailed observation of the current
heap size at any time of execution ($m(t)$). The maximum total size of the heap
($m(t_{max})$) has been determined and an average memory consumption
$\overline{M}$ has been calculated as the integral of the current
memory consumption $m(t)$ over the time of execution normalized by execution time.

\begin{equation}
\overline{M} = \int_{t\_start}^{t\_end}\frac{m(t)}{{t\_end}-{t\_start}}\,dt ~ ~
[byte]
\end{equation}

\begin{figure}[ht]
    \begin{minipage}[c]{0.5\textwidth}
    \includegraphics[width=\textwidth]{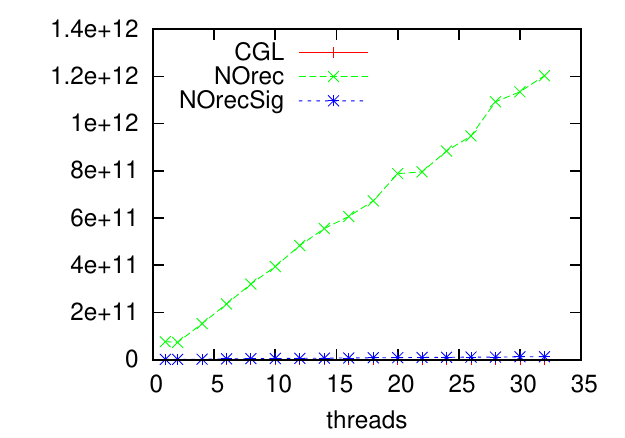}
	\caption{opactiy: $m(t_{max})$ [byte]}
	\label{fig:memmax.opactiy}
	\end{minipage}
\quad
    \begin{minipage}[c]{0.5\textwidth}
    \includegraphics[width=\textwidth]{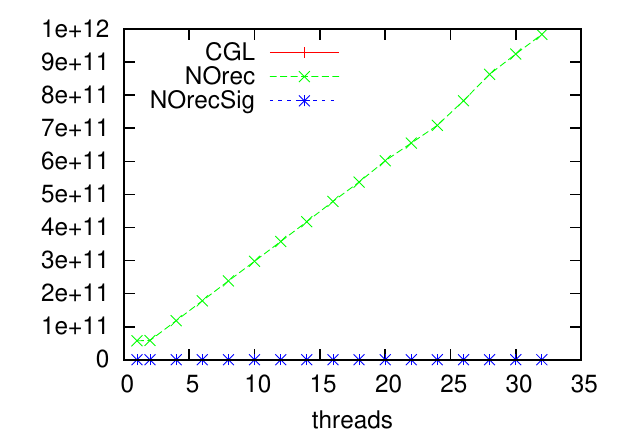}
	\caption{opactiy: $\overline{M}$ [byte]}
	\label{fig:memavg.opactiy}
 	\end{minipage}
\end{figure}

\begin{figure}[ht]
    \begin{minipage}[c]{0.5\textwidth}
    \includegraphics[width=\textwidth]{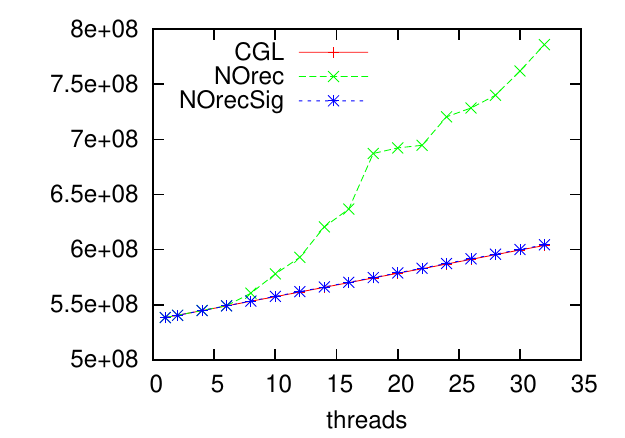}
	\caption{intruder: $m(t_{max})$ [byte]}
	\label{fig:memmax.intruder}
	\end{minipage}
\quad
    \begin{minipage}[c]{0.5\textwidth}
    \includegraphics[width=\textwidth]{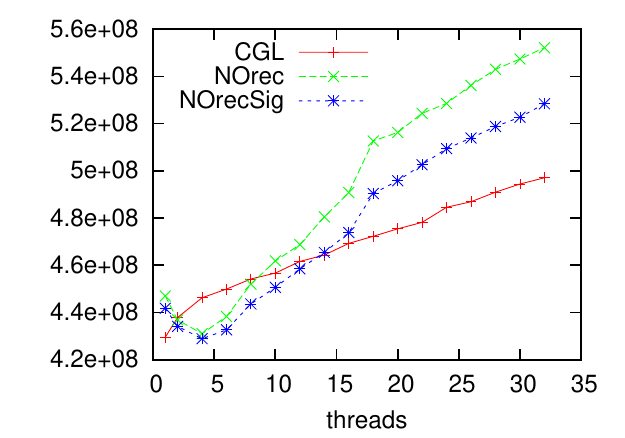}
	\caption{intruder: $\overline{M}$ [byte]}
	\label{fig:memavg.intruder}
 	\end{minipage}
\end{figure}

Memory consumption of \texttt{NOrecSig} was in most cases significantly lower and in some
cases even similar. The worst case in maximum and average was reached with our
benchmark (Figures \ref{fig:memmax.opactiy} and \ref{fig:memavg.opactiy}).
The difference of 1.2TB\footnote{Memory is not accessed and thus can grow over
the actual capacity of the hardware} in maximum memory consumption illustrates
the danger of resource exhaustion. The intruder benchmark (Figures
\ref{fig:memmax.intruder} and \ref{fig:memavg.intruder}) shows another case
where maximum and average memory consumption of \texttt{NOrec} are just moderately higher than that of \texttt{NOrecSig}.

\section{Conclusion}
\label{sec:conclusion}
The review of techniques to deal with memory access violations of doomed
transactions due to memory reclamation after privatization revealed remaining
issues with incremental validation and epoch-based memory reclamation.
Inspired by sandboxing, the approach proposed here is to use incremental
validation and handle generated segmentation fault signals to recover from
inconsistent states of the transaction. This method solves all memory access
violations occurring from memory reclamations inside and outside of transactions.
A comparison of a prototype to NOrec has proven, that it slightly
improves the response time of memory management functions and effectively
decreases memory consumption. That way it improves the opacity of incremental validation
because it prevents unexpected memory exhaustion to occur.  A future comparison
to sandboxing will clarify their relationship in terms of throughput and
scalability.

\bibliographystyle{splncs}
\bibliography{ARCS2014-Opacity-of-MM-in-TM}

\end{document}